\documentstyle[fleqn]{tp}

\input psfig

\def\puncspace{\ifmmode\,\else{\ifcat.\C{\if.\C\else\if,\C\else\if?\C\else%
\if:\C\else\if;\C\else\if-\C\else\if)\C\else\if/\C\else\if]\C\else\if'\C%
\else\space\fi\fi\fi\fi\fi\fi\fi\fi\fi\fi}%
\else\if\empty\C\else\if\space\C\else\space\fi\fi\fi}\fi}
\def\SP{\let\\=\empty\futurelet\C\puncspace}

\def\iras{{\it IRAS}\SP}

\def\kms{km~s$^{-1}$\SP}

\def\h1{$h^{-1}$\SP}

\def\lsim{~\rlap{$<$}{\lower 1.0ex\hbox{$\sim$}}}
\def\gsim{~\rlap{$>$}{\lower 1.0ex\hbox{$\sim$}}}
\def\void#1{{}}
\def\etal{{\it et al.\/}\ }

\def\eg{{\it e.g.\/}\rm,\ }

\begin{document}

\twocolumn[
\title{Galaxy Redshift Surveys: 20 years later}
\author{Luiz da Costa$^1$\\
{\it $^1$ESO, Karl-Schwarzschildstr. 2, D--85748 Garching}}
\vspace*{16pt}   

ABSTRACT.\ This year marks the 20th anniversary of the effective
beginning of large, systematic redshift surveys of galaxies. These
surveys have had a major impact on observational cosmology and on our
current understanding of large-scale structures in the Universe. To
celebrate this remarkable period some landmark observational results
are reviewed and our current understanding of LSS is summarized.
Although enormous progress has been achieved in mapping the galaxy
distribution to moderate redshifts, many of the questions posed over a
quarter of century ago have not yet been convincingly answered and
must await the completion of new large solid angle surveys such as 2dF
and SDSS.  On the other hand, unexpected advances have been made at
very high redshifts. After years of searching, well-defined samples of
extremely distant galaxies are now available, redshifts are being
routinely measured and large programs are planned for the 8-m class
telescopes. These ongoing or planned surveys of the nearby and distant
Universes promise to provide, within a few years, an extraordinary
view of the evolution of galaxies and structures from lookback times
approaching $\sim$90\% of the age of the Universe to the present
epoch.

\endabstract]

\markboth{Luiz da Costa}{Galaxy Redshift Surveys: 20 years later}

\small

\section{Introduction}

Redshift surveys of galaxies have been for the past two decades one of
the most useful tools available for cosmological studies.  Since the
pioneering work of Gregory and Thompson (1978) and Sandage (1978),
among others, progress has been enormous, with the pace being dictated
primarily by advances in technology. Arguably one of the most
significant early milestones in the field was the start of the Center
for Astrophysics Redshift Survey in 1978, one of the first surveys to
replace photographic plates by significantly more efficient detectors,
making large scale surveys possible. Similar systems soon became
available at several observatories and the number of surveys, using
different samples and observing strategies, flourished in the 80s.
New developments in the 90s further increased the data gathering power
of surveys making possible to probe much larger volumes. Combined
these surveys have provided a wealth of information regarding the
properties of galaxies, systems of galaxies and the nature of
large-scale structure (LSS) as traced by galaxies.

Despite the enormous effort a number of outstanding cosmological
questions remain unanswered pointing out the need for even larger
surveys. Projects like the Anglo-Australian 2dF and the Sloan Digital
Sky Survey (SDSS)  represent the beginning of a new era in the field and
promise to provide clear answers to these questions. Another milestone
are the ongoing (Steidel, this proceeding) and the planned redshift
surveys of the high-z Universe on Keck and VLT which will provide a
better understanding of the nature of distant galaxies, their clustering
properties and insight into early galaxy and structure evolution.

Over the years the goal of most galaxy redshift surveys have remained
essentially the same, namely to obtain redshifts for complete samples
over a sufficiently large volume to: 1) study the nature of large
scale structure; 2) measure the cosmological density parameter
$\Omega$ from dynamical measurements on small and large scales; 3)
compare observed galaxy distribution to predictions based on $N-$body
simulations in an attempt to discriminate among competing theoretical
models; 4) compare the galaxy distribution to the mass distribution as
recovered, for instance, from the peculiar velocity field of galaxies;
and 5) study galaxy biasing on small and large scales. Even though
galaxy redshift surveys alone provide only limited information about
the underlying mass fluctuations, they will continue to be essential
for probing galaxy biasing and evolution models, complementing the
information from probes of the mass distribution such as cosmic flows,
gravitational lensing and cosmic microwave background radiation.

The literature on large-scale structure has grown tremendously and a
comprehensive review on the subject would be well beyond the scope of
this presentation and can be found elsewhere (\eg Giovanelli \& Haynes
1991, Strauss \& Willick 1995). Instead, the aim of this review is to
illustrate how our picture of the Universe has evolved over the past
two decades with the completion of various surveys targeting
different redshift intervals. Also reviewed are the results of
quantitative analyses carried out with redshift data to describe
galaxy clustering as well as the properties of the galaxy population
as a whole.

\section{Background}

In this section, a brief review is presented of the redshift surveys
that have shaped our current understanding of the three-dimensional
distribution of galaxies.  Over the years the picture has evolved
dramatically as data flowed in.  The number of galaxies with redshift
measurements has grown by roughly a factor 100 from the mere $\lsim
2000$ available in 1978. This number is expected to grow even faster
now as new data from surveys such as 2dF and SDSS and other
large surveys at high-redshift  become available.

\subsection {Low-z}

Three-dimensional redshift surveys, which densely sample the local
galaxy distribution, are essential to characterize the properties of
galaxies and the nature of the large-scale structures at the present
epoch. Dense sampling is critical for studying the morphology of large
scale structures, while the number of galaxies and surveyed volume are
necessary for detailed statistical analyses. It is important to
emphasize that the nearby Universe is in some respects unique. For
instance, only nearby can one expect to use the peculiar velocity
field of galaxies to map the mass distribution in the framework of the
gravitational instability paradigm. Comparison between the galaxy and
the reconstructed mass distributions provides a valuable probe of the
relation between galaxies and the underlying dark matter, at least on
large scales (Dekel 1994, da Costa \etal 1996). Furthermore, linear
theory gives a relation between galaxy density and peculiar velocity,
which can be used to derive a velocity field from all-sky redshift
surveys. The derived velocity field can be compared to that observed
to measure $\beta = \Omega^{0.6}/b$, where $\Omega$ is the
cosmological density parameter and $b$ is the linear biasing factor
(\eg Davis, Nusser \& Willick 1996, da Costa \etal 1998).

\void{
\begin{figure*}
\centering\mbox{\psfig{figure=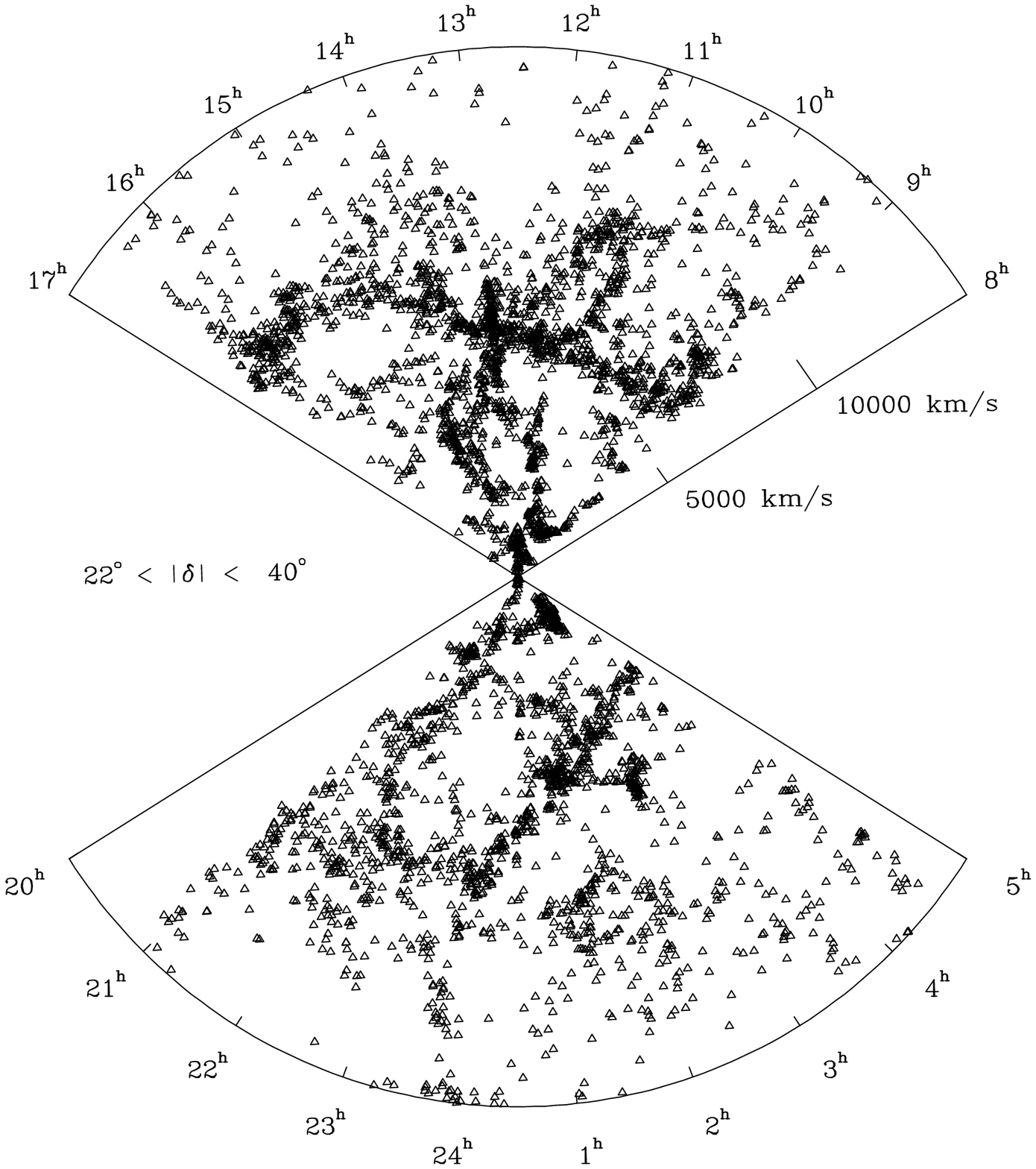,height=6cm}}
\vspace*{0.25cm}
\caption[]{Two-dimensional distribution of galaxies with complete
redshift information from CfA2 and SSRS2.}
\label{fig:projdist}
\end{figure*}
}

Until the mid-70s all the information available to cosmologists was
the projected galaxy distribution.  Although it was apparent that the
distribution was far from homogeneous nothing was known about the
reality of the observed structures in three-dimensions. The pioneering
pencil-beam surveys of nearby clusters and surroundings (\eg Gregory
\& Thompson 1978) provided the first hint that the galaxy distribution
was irregular, motivating wider-angle surveys. Such an attempt was
first carried out by Sandage (1978). However, the sample, which had a
median radial velocity of 1500 \kms, was too shallow to probe the
nature of the large-scale structures.  The CfA Redshift Survey (Davis
\etal 1982) was the first wide-angle survey to reach beyond the Local
Supercluster. It provided strong evidence that the galaxy distribution
was far from homogeneous, showing instead a complex topology made up
of large empty regions and filaments.  However, the structures were
poorly defined because of the sparseness of the sample. The picture
that emerged was considerably different from that envisioned just a
few years earlier, in which clusters were believed to be rare,
isolated regions of high density in an otherwise uniform
background. In parallel, the KOSS survey (Kirshner \etal 1981), using
a series of pencil-beam surveys, identified a large empty region with
an estimated size of 6000\kms. Pressing the observations to fainter
magnitudes, the survey of a thin $6^\circ$ wide slice was completed
(de Lapparent, Geller \& Huchra 1986) showing that these empty regions
were bound by remarkably sharp and coherent structures with scales
comparable to the linear size of the survey ($\sim 100$\h1 Mpc).
However,the slice-like geometry of the survey did not allow one to
differentiate between two-dimensional structures or one-dimensional
filaments. Further evidence of large coherent structures came from the
HI Arecibo Survey of the Perseus-Pisces region (\eg Giovanelli \&
Haynes 1991).  More convincing proof that these coherent structures
were not filaments but two-dimensional sheets came from the Southern
Sky Redshift Survey (SSRS, da Costa \etal 1988). The SSRS was designed
to extend the CfA survey to the southern hemisphere in order to: 1)
obtain a unobstructed view of the LSS by avoiding the Virgo cluster;
2) to produce the first all-sky sample to moderate depth; 3) to test
the reproducibility of different statistics employed in the analysis
of the CfA data. By construction the southern sample was required to
have the same surface density as in the north.  Since there were no
prominent nearby clusters this requirement led to a slightly deeper
survey which allowed for the detection of the Southern Wall, a
coherent, thin structure seen over the entire declination range probed
by the SSRS.

The need for better sampling of the structures motivated the extension
of the CfA and SSRS surveys to fainter magnitude limits.  The CfA2
Redshift Survey (Geller \& Huchra 1989) soon produced the striking map
of the full extent of the structure seen in their earlier slice-survey
-- the Great~Wall -- a spectacular example of a thin two-dimensional
structure containing several rich clusters.  Following suit, the SSRS2
(da Costa \etal 1994a) confirmed that the Great Wall was not a rare
event, even though unique is some aspects, but that large voids and
walls are in fact common features of the galaxy distribution.

Combined, CfA2 and SSRS2 now cover over 30\% of the sky to the same
depth. The sample consists of over 20,000 galaxies, a remarkable
progress relative to the first generation of wide-angle
surveys. Together, the CfA2 and SSRS2 surveys
provide a unique and, so far, unmatched database which combines dense
sampling with an almost complete three-dimensional view of the
present-day galaxy distribution. They extend out to a moderate depth
($cz$ \lsim 15,000 \kms), which allows one to probe a linear scale of
the order of 300 \h1 Mpc.

\begin{figure*}
\centering\mbox{\psfig{figure=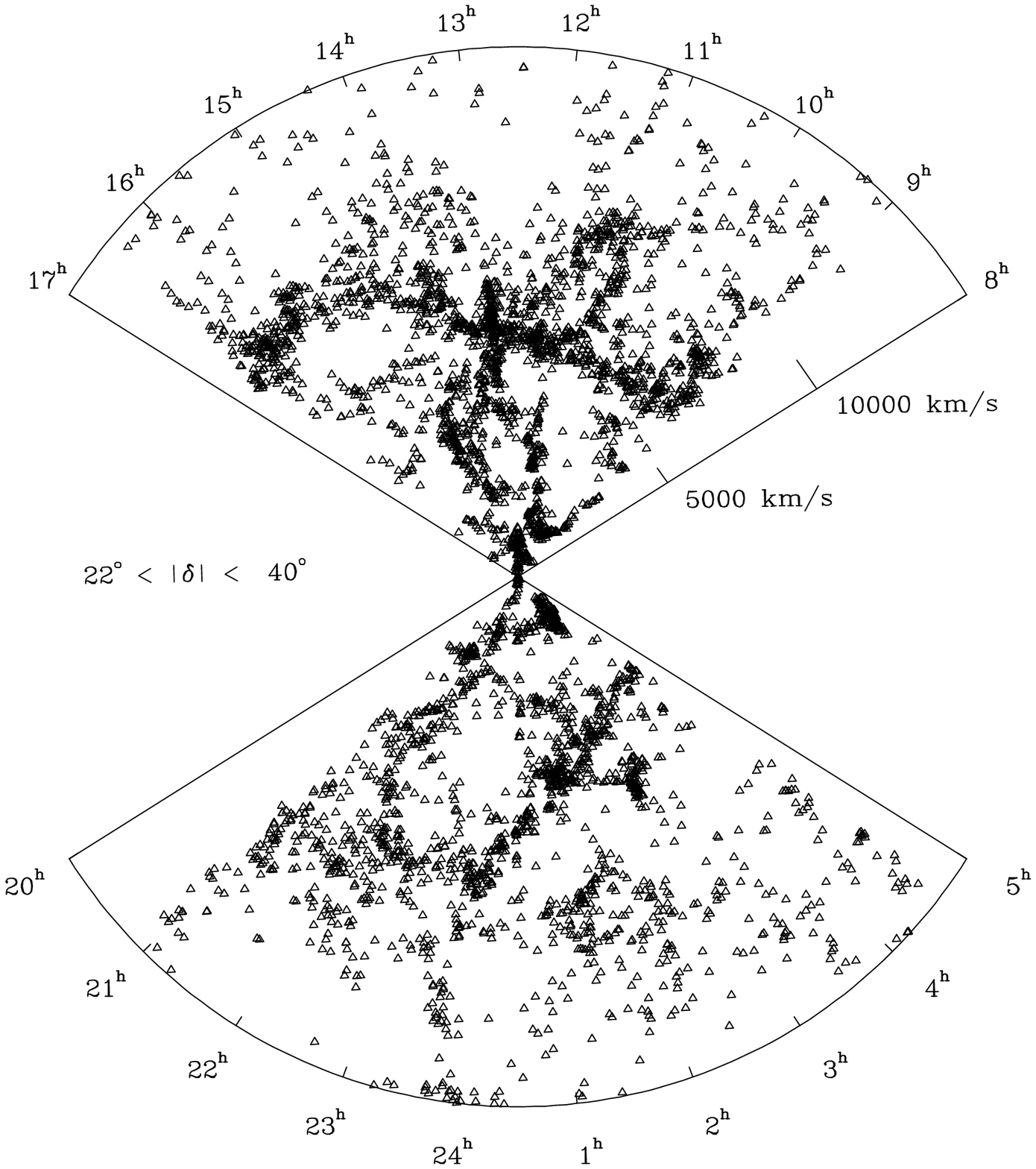,height=6cm}}
\vspace*{0.25cm}
\caption[]{Redshift versus right ascension diagram for galaxies
brighter $m_B \leq 15.5$ and within a $10^\circ$ wedge taken from the
combined CfA2-SSRS2 sample.}
\label{fig:crosssection}
\end{figure*}

Figure~\ref{fig:crosssection} shows a cross-section of the local galaxy distribution obtained
by combining the CfA2 and SSRS2.  From the redshift maps alone one finds
that large coherent structures appear to be a common feature of the
galaxy distribution. Walls and voids, 5000 \kms in diameter, are seen in
every region large enough to contain them. The qualitative picture that
emerges is one in which the galaxy distribution consists of a
volume-filling network of voids.

Unfortunately, the scope of optical surveys is limited to relative
high galactic latitudes because of the zone of avoidance. Therefore,
to extend the sky coverage one must resort to infrared-selected
samples. Examples of redshift surveys based on \iras galaxies include
the 1.2 Jy \iras Survey (Fisher \etal 1995), QDOT (\eg Kaiser \etal
1991) and more recently PSCz (Saunders \etal 1998). Although
considerably more sparse than their optical counterparts, the main
advantage of \iras galaxy redshift surveys is the uniform and
unmatched all-sky coverage. Full sky-coverage greatly simplifies
statistical analyses (\eg power-spectrum analysis, counts-in-cells),
bypassing some of the problems associated with edge effects and survey
geometry. Equally important is the uniformity of the parent sample,
which eliminates some of the uncertainties that have plagued the
nearby optical surveys. But, above all, the main contribution of
redshift surveys of \iras galaxies is the fact that only from them can
one compute a reliable peculiar velocity field as predicted from the
galaxy density field, a key element in understanding the dynamics of
the local Universe.

Because of the unexpected large size of the structures observed
nearby, it became essential to extend the surveys to greater
depths. To achieve this goal in a reasonable amount of time, the
Stromlo-APM survey (Loveday \etal 1992) used a sparse-sampling
technique advocated by Kaiser (1986), ideal for low order statistics,
measuring redshifts for about 1800 galaxies randomly drawn at a rate
of 1 in 20 from a complete magnitude-limited catalog selected from the
APM Galaxy Survey (Maddox \etal 1990a). The survey probes a depth of
$\sim$ 200\h1 Mpc, sampling a volume about five times that of the
CfA2-SSRS2, at the expense of small-scale information.  The data were
used to measure the luminosity function of galaxies and their
clustering properties on large scale,with the large volume allowing
for the sampling a large number of different structures. Analysis of
the radial density variation also showed no evidence of a large local
void, one of the proposed explanations for the strong variation of the
APM galaxy number counts at the bright end (Maddox \etal 1990b). 

All the previous surveys were carried out observing a galaxy at a
time.  A major progress in redshift surveys came about with the
multiplexing capability of multi-object spectrographs. An outstanding
example of the benefits of the combination of fiber-fed spectrographs
and wide-field telescopes is the recently completed Las Campanas
Redshift Survey (LCRS, Schectman \etal 1996) carried out on the 2.5m
du Pont telescope at Las Campanas. The LCRS contains redshifts for
over 25,000 R-selected galaxies covering 0.2 steradians in six strips,
each 1.5$^\circ \times 80^\circ$, in the south and north Galactic
caps. The median redshift of the survey is $z\sim0.1$. Although
probing a much larger volume, about five times larger than the
combined CfA2-SSRS2 survey, inspection of the redshift maps supports
the picture that the galaxy distribution consists of a closely-packed
network of voids $\sim$ 5000 \kms in diameter bounded by thin, large
walls, with no strong evidence of inhomogeneities on larger scales.

More recently, other surveys to comparable depth to the LCRS, but
adopting different selection criteria and observing strategies, have
been completed: the Century Survey (CS, Geller \etal 1997) with about
1800 galaxies covering 0.03 steradians and the ESO Project Slice (EPS,
Vettolani \etal 1997) with about 3,300 galaxies covering 0.008
steradians. Again the large-scale features are qualitatively similar
to those seen in earlier surveys. However, both surveys claim to find
evidence of inhomogeneities, such as the Corona Borealis supercluster,
on a scale of $\sim$100\h1 Mpc.

Further progress in this range of redshifts will have to await the
completion of 2dF and SDSS which will measure of the order of a million
redshifts, providing a complete and unprecedented wide-angle coverage of
the galaxy distribution to a depth of about 300\h1Mpc. The impact that
these surveys will have can already be appreciated from the preliminary
results of the 2dF survey (Maddox, this proceedings).

\subsection {Intermediate-z}

Hints for power on very large scales were first detected by the deep
BEKS (Broadhurst \etal 1988) pencil-beam survey extending to $z\sim
0.5$ in the direction of the Galactic poles. The survey used a
collection of narrow probes to map the galaxy distribution over a
linear scale of about 2,000\h1 Mpc.  The observed distribution shows a
remarkable regularity exhibiting an alternation of peaks and voids
with a typical scale of 128\h1 Mpc. However, follow-up observations in
other directions not only do not confirm this regularity but detect
power on smaller scales ($\sim$60\h1 Mpc), in agreement with the
results from nearby surveys. In order to verify these claims of large
scale power, the ESO-Sculptor redshift survey (ESS, Bellanger \& de
Lapparent 1995) was designed to have a transverse dimension larger
than the galaxy correlation length at the median redshift of the
survey ($z \sim 0.3$) to assure that the detected structures are not
artifacts caused by small-scale clustering, one of the main criticisms
to the original interpretation of BEKS results.  The ESS provided the
first detailed map of the galaxy distribution in the redshift interval
$0.1<z<0.5$, which confirmed the existence of voids bounded by thin
structures over the entire redshift interval. More importantly, the
ESS confirmed that the voids have a typical size of $\sim$ 60\h1 Mpc,
finding no evidence for periodic structures on scales $\sim$ 130\h1
Mpc.

Other deep surveys, extending to even larger redshifts ($z \lsim 1$),
have also been completed but have focused primarily on the direct
study of the evolution of the luminosity function, star formation and
clustering. Among them are: 1) the Autofib Redshift Survey (Ellis
\etal 1996) which combines several pencil-beam surveys of
magnitude-limited samples (1700 galaxies) spanning a wide range in
apparent magnitude down to $b_j$ =24, and reaching $z\sim0.75$; 2) the
Canada-France Redshift Survey (CFRS, Lilly \etal 1995) containing some
600 galaxies brighter than $I_{AB}$=22.5, with a median redshift of
$z\sim0.56$, and covering an effective solid angle of
112~arcmin$^2$. These data have provided one of the first secure
evidence of a physical association of galaxies at $z \sim 1$ (Le
F\`evre \etal 1994); 3) The CNOC2 survey (Carlberg \etal 1998) which
presently contains about 5,000 galaxies with $R<21.5$ to $z \sim 0.6$,
over a total area of 1.5 square degrees.

Like in the past, these first results have motivated different groups
to plan ambitious surveys using 8-10m class telescopes such as Keck
(DEEP) and VLT (VIMOS) to observe large samples of galaxies ($\sim
10^5$), probing scales of $\sim$ 100\h1 Mpc at $z\sim1$ (\eg Le
F\`evre \etal 1996a). Completion of these surveys will allow one to
put together a coherent and self-consistent picture of galaxy and
clustering evolution from $z \sim 1$ to the present.

\subsection {High-z: the new frontier}

Perhaps one of the most exciting developments in recent years has been
the discovery of a population of luminous, star-forming galaxies at
$z\sim3$, using a color criteria sensitive to the presence of the
Lyman continuum break (Steidel \& Hamilton 1993). Currently, the
spatial distribution of these Lyman break (U-dropout) objects is being
investigated and spectroscopic redshifts have been obtained for about
500 galaxies (Steidel, this proceedings). Preliminary results have led
to the discovery of a large structure at $z \sim 3.1$, which could
indicate that well developed, large-scale structures exist at even
these high redshifts (Steidel \etal 1998).

Another important development is the construction of near-IR
spectrographs (\eg NIRMOS-VLT) to measure galaxy redshifts in the
interval $1<z<3$, where most of the spectral features lie outside the
optical window. This will allow to bridge the gap that currently
exists between the low-z ($z\lsim1$) and high-z ($z\gsim3$) domains.

\begin{figure*}
\centering\mbox{\psfig{figure=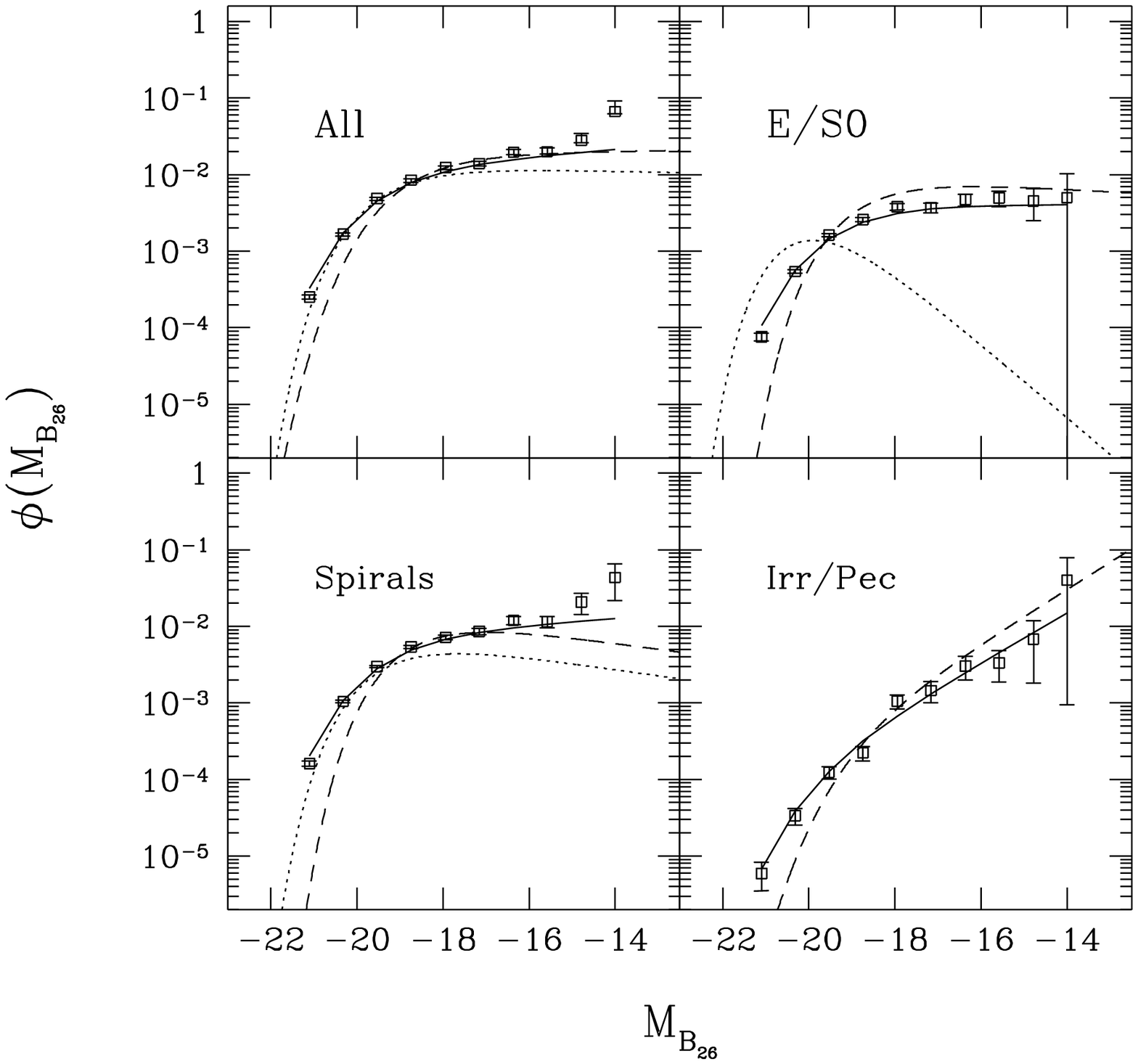,height=8cm}}
\vspace*{0.25cm}
\caption[]{A comparison of recent measurements of the local LF divided
by morphological type.  Solid lines and open squares represent the
SSRS2; dashed lines represent the CfA Survey and dotted lines
represent the Stromlo-APM (for details see Marzke \etal 1998). }
\label{fig:lf}
\end{figure*}

\section{Quantitative Results: Highlights}

Considerable progress has also been made in measuring the properties
of LSS and galaxies using the data available from the redshift surveys
described above. In general, the results from different surveys are
consistent, even though some discrepancies still persist and some
basic questions remain unanswered. In this section, our current
understanding is reviewed by highlighting some of the main
quantitative results.

\subsection{Luminosity Function}

Perhaps one of the most basic statistic that can be measured from
redshift surveys is the luminosity function (LF). It not only provides
information about the galaxy population but it is a key ingredient in
the analysis of magnitude-limited samples. The importance of
determining the local LF is that its overall normalization and
faint-end slope directly impact the interpretation of the excess
observed in the faint number counts and the amount of evolution
required to explain them.  Furthermore, the local LF has been used to
calibrate or to verify the consistency of semi-analytical galaxy
evolution models, which have become a powerful tool for detailed
comparisons between data and theory (\eg Kauffmann, White \&
Guiderdoni 1993, Lacey \etal 1993).  However, despite the innumerous
estimates of the local LF, there still is considerable debate over its
shape and normalization. The nature of the problem is reviewed below
using results from the most recent surveys.

The local LF has been independently computed for the CfA2 (Marzke
\etal 1994a) and SSRS2 (da Costa \etal 1994a, Marzke \etal 1998) north
and south sub-samples, which altogether probe four different regions
of the sky. Comparison of these LFs shows that the shapes are in
relative good agreement, especially at the faint-end.  However, the
derived normalization of the CfA2 north LF is significantly higher
than the rest, suggesting a mean galaxy density a factor of two larger
in that region.  By contrast, the LF measured for the SSRS2 south and
north are essentially identical, presenting very similar shapes and
normalizations, even though they probe distinct volumes and largely
independent structures. Their normalization is also consistent with
that derived for CfA2 south. There are two possible interpretations
for the observed discrepancy: 1) there are significant fluctuations of
the galaxy distribution on scales of $\sim$ 100\h1 Mpc; 2) there are
systematic errors in the magnitude-scale, in particular those given in
the Zwicky catalog from which the CfA2 sample is drawn.

The same disagreement is seen when comparing the LF of more distant
samples (Ellis \etal 1996, Lin \etal 1996a, Zucca \etal 1997, Geller
\etal 1997). In general, the derived luminosity functions fall into
two broad categories - those with high (Autofib, CfA2, CS, ESP) and
low normalizations (SSRS2, Stromlo-APM, LCRS). Again, with the
exception of the CfA2 (at the bright end) and LCRS (at the faint end)
the shapes are, by and large, very similar.  These results are
puzzling since there is no clear correlation between the samples used
and the direction in space or the way the parent catalogs for these
samples were created.  Possible explanations for the conflicting
results include: the existence of a large underdense region in the
local Universe, an underestimate of a population of low luminosity
galaxies nearby or a rapid evolution of the blue luminosity function
at low redshifts ($z \sim 0.1$).

The local LF has also been examined as a function of morphology
(Figure~\ref{fig:lf}) and color using the CfA2 (morphology) and SSRS2 (morphology
and color) samples.  Analyzes of these samples show that even locally
one observes an excess of blue galaxies at faint magnitudes. It is
estimated that the faint-end slope of blue galaxies is $\alpha
\lsim$~-1.3 (Marzke \& da Costa 1997). In addition, using the complete
morphological information available for the SSRS2, one finds that
early and late-type galaxies have very similar, flat LFs (Marzke \etal
1998), while the irregular/peculiar galaxy LF is very steep ($\alpha
\sim -1.81$). These results are in good agreement with earlier
findings based on the CfA data (Marzke \etal 1994b) but in clear
disagreement with the results of Loveday \etal (1995), probably
because of inadequacies in the identification of ellipticals at faint
magnitudes. Similar studies are currently underway at moderate
redshifts.

A clear resolution of some of the problems mentioned above will have
to await the completion of SDSS which will provide a homogeneous,
multi-color photometric data set of the northern sky with complete
redshift information.

Recent surveys such as Autofib and CFRS, and now CNOC2. with an
extended redshift baseline, have provided for the first time, the
means to directly study the evolution of the luminosity function. The
CFRS sample has been subdivided into several redshift bins and into
two colors. Analysis of these subsamples shows that the redder
galaxies exhibit remarkably little evolution, while strong evolution
is observed for the bluer galaxies. It is important to note that this
evolution is independent of the normalization of the "local" LF since
it is determined from the sample itself. Strong evolution of blue
galaxies has also been observed in the B-selected sample of Autofib.

\subsection{Galaxy Power-Spectrum}

The power-spectrum (PS) of galaxies in redshift space has been
computed for a number of optical (\eg Park, Gott \& da Costa 1992,
Park \etal 1994, da Costa \etal 1994b, Lin \etal 1996b) and infrared
surveys (Fisher \etal 1993). The redshift-space PS estimates roughly
follow a power-law $P(k)\propto k^n$ with a slope ranging from $n \sim
-2$ on small scales ($\lambda \lsim$30\h1 Mpc) to $n \sim -1.1$ on
intermediate scales (30\h1 $<\lambda<$120\h1 Mpc). For nearby samples,
such as the combined CfA2-SSRS2, one finds that the PS continues to
rise on scales up to $\sim$ 200\h1 Mpc.  This result has been
confirmed by similar analysis of other optical and infrared-selected
samples, all showing essentially the same shape, while differences in
the amplitude can be ascribed to the relative bias between optically
and infrared-selected galaxies or between galaxies of different
luminosities. These earlier results have been confirmed by the PS
computed from the LCRS which shows good agreement with previous
calculations on scales $\lsim 100$ \h1 Mpc. On larger scales, the LCRS
PS shows a change in slope and strongly suggests that it has detected
the turnover. A good fit for the observed PS in redshift space,
satisfying the constraints implied by COBE, can be obtained with a
open or flat nonzero cosmological constant CDM model with a shape
parameter $\Gamma=\Omega h$=0.2 with no bias. However, several other
models are equally viable (da Costa \etal 1994b, Lin \etal 1996b).

\begin{figure*} 
\hspace*{6cm} \centering\mbox{\psfig{figure=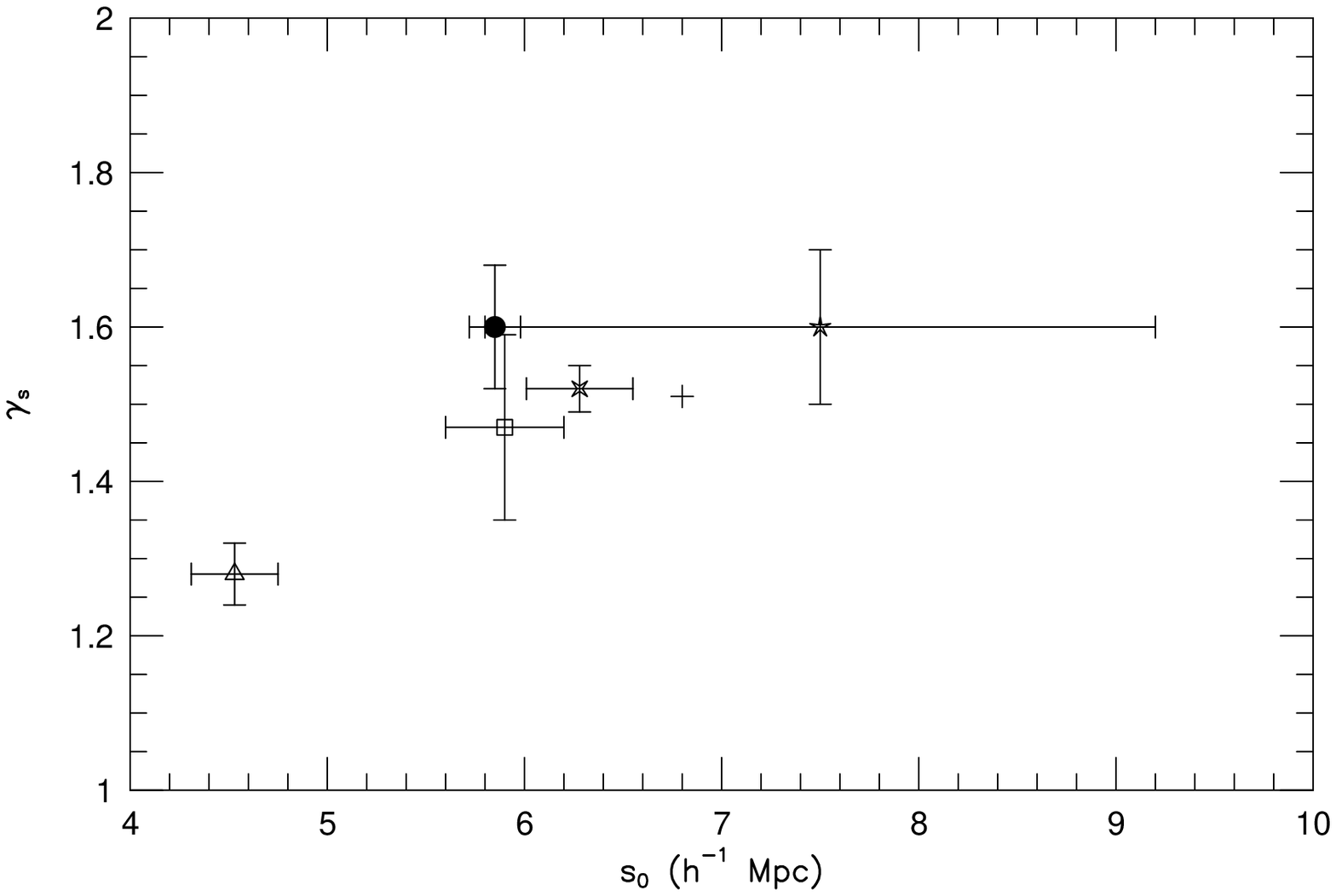,height=6cm,angle=270}}
\vspace*{0.25cm} \caption[]{Comparison of the correlation length and
slope of the two-point correlation function in redshift space computed
for different optical and IR samples. The samples considered are:
CfA2-slice (filled triangle), SSRS2 (filled square), LCRS (star),
Stromlo-APM (open squares), ORS (cross), IRAS (open triangles) (for
details see Willmer, da Costa \& Pellegrini 1998).} 
\label{fig:correlation}
\end{figure*}

\subsection {Correlation Function}

The two-point correlation function, formally equivalent to the PS, has
been the most widely used statistics to quantify galaxy
clustering. Analyzes of magnitude-limited samples have led to
consistent results between nearby surveys (\eg SSRS2) and those
probing volumes more than five times larger and adopting different
survey geometry and sampling strategies (\eg Stromlo-APM, LCRS). A
summary of these results is shown in Figure~\ref{fig:correlation}, where the redshift
correlation length $s$ and the slope $\gamma$ of the best-fit power
law derived from different samples are compared (\eg Willmer, da Costa
\& Pellegrini 1998, and references therein). As can be seen there is a
remarkable agreement among the optical surveys, except for the CfA2
where peculiar motions near the Great Wall are important. In
particular, note that the good agreement between the values found for
a wide range of volumes is in marked contrast with what would be
expected if the galaxy distribution were a fractal. In some cases the
relatively small differences between redshift and real space
correlations on intermediate scales ($\sim$ 10\h1 Mpc) immediately
suggest, as in the case of the SSRS2, a low value of $\beta =
\Omega^{0.6}/b < 1$ (\eg Willmer, da Costa \& Pellegrini 1998).

While galaxy clustering at the present-epoch seems to be well
quantified, at least for low-order statistics, work is now concentrated
in measuring its evolution and interpreting the results within the
hierarchical clustering framework. Until recently such studies were
hindered by large uncertainties as they had to rely on the observed
two-point angular correlation function and models for the clustering
evolution and redshift distribution. However, the availability of an
increasing number of samples (CFRS, CNOC2) spanning a large redshift
baseline, will soon provide the means to directly measure the
evolution of the correlation function and disentangle the effects of
cosmology and galaxy evolution (\eg Kauffmann \etal 1998). Preliminary
results from CFRS indicate a strong evolution of the clustering
amplitude with redshift up to $z \sim 0.6$ (Le F\`evre \etal 1996b).
This is at variance with more recent results of Carlberg \etal (1998)
based on the CNOC2 data, who find a much weaker evolution. Results
from very high redshift surveys are also becoming available (Steidel,
this proceedings) and the time evolution of clustering as a function
of galaxy properties is within reach.

\subsection {Higher-order Statistics}

Given the complexity of the observed large-scale structure, a complete
statistical description of the galaxy distribution requires the use of
high-order statistics.  To investigate high-order correlations,
counts-in-cells have been used to compute the count probability
distribution function $P(N,V)$, from which the Void Probability
Function (VPF), $P(0,V)$, and the normalized skewness $S_3$ and
kurtosis $S_4$ have been derived. These statistics have been used to
test the hierarchical relations and to compare data to simulations
using optical and infrared-selected samples with complete redshift
information (Vogeley \etal 1991, Lachieze-Rey, da Costa \&
Maurogordato 1992, Bouchet \etal 1993, Benoist \etal 1998). From the
moments of the counts distribution and from the scaling of the VPF one
finds that the galaxy distribution satisfies the scaling relations
predicted by second-order perturbation theory well into the non-linear
regime. However, high-order statistics, such as VPF, have not proven
to be good discriminants of different cosmological models. Instead,
preliminary results suggest that high order moments may be best used
to constrain galaxy biasing models, especially for large redshift
samples expected from 2dF and SDSS.

\subsection {Small-scale Velocity Field}

Redshift surveys can provide statistical estimates of deviations of
the Hubble flow on small scales, without requiring direct distance
measurements of individual galaxies. As discussed by Davis \& Peebles
(1983) this can be done by examining the correlation function $\xi$,
as a function of the projected separation $r_p$ and redshift
separation $\pi$ of pairs. Deviations of $\xi (r_p,\pi)$ from
concentric circles are due to redshift distortions, which provide
information on the distribution function of relative peculiar
velocities of galaxy pairs. On large scales, linear theory relates the
first moment of this distribution to the density parameter $\Omega$
and the linear bias parameter $b$. On small scales, the cosmic virial
theorem connects the second moment to these parameters.

Analysis of redshift distortions observed the 1.2 Jy \iras Survey lead
to estimates of $\Omega/b \sim$ 0.4, from the cosmic virial theorem,
and $\beta = \Omega^{0.6}/b = 0.45$ on scales $\sim$ 10\h1 Mpc (Fisher
\etal 1994).  Assuming that the relative bias between optical and
\iras galaxies is $b_o/b_I \sim 1.5$ this result implies that
$\sigma_8\Omega^{0.6} \sim 0.3$, where $\sigma_8$ is the rms mass
fluctuation within a sphere 8\h1 Mpc in radius.  Unfortunately, both
estimates suffer from either large systematic errors or large cosmic
variance, due to the limited number of independent structures sampled
by the nearby surveys. This has been vividly illustrated by the large
sample-to-sample variations of the relative velocity dispersion
between pairs derived from the combined CfA2-SSRS2 sample (Marzke
\etal 1995).  The finding that this quantity shows strong
sample-to-sample variations indicates that it is poorly determined
within the volume surveyed, being dominated by the shot-noise
contribution of clusters.  One is forced to conclude that at the
present time the small-scale velocity field is not a powerful
discriminant among competing cosmological models. Even though new
statistics have recently been proposed to overcome the effects of a
pair-weighted statistic (Davis, Miller \& White 1997, Strauss,
Ostriker \& Cen 1998), it is clear that for robust measurements
considerably larger volumes, sampling a fair number of clusters of
different richness, are required. This will certainly be possible with
the next generation of surveys. It is worth pointing out that the
estimates of $\beta$ on small scales are consistent with the most
recent estimates of this parameter from cosmic flows (\eg da Costa
\etal 1998).

\begin{figure*}
\centering\mbox{\psfig{figure=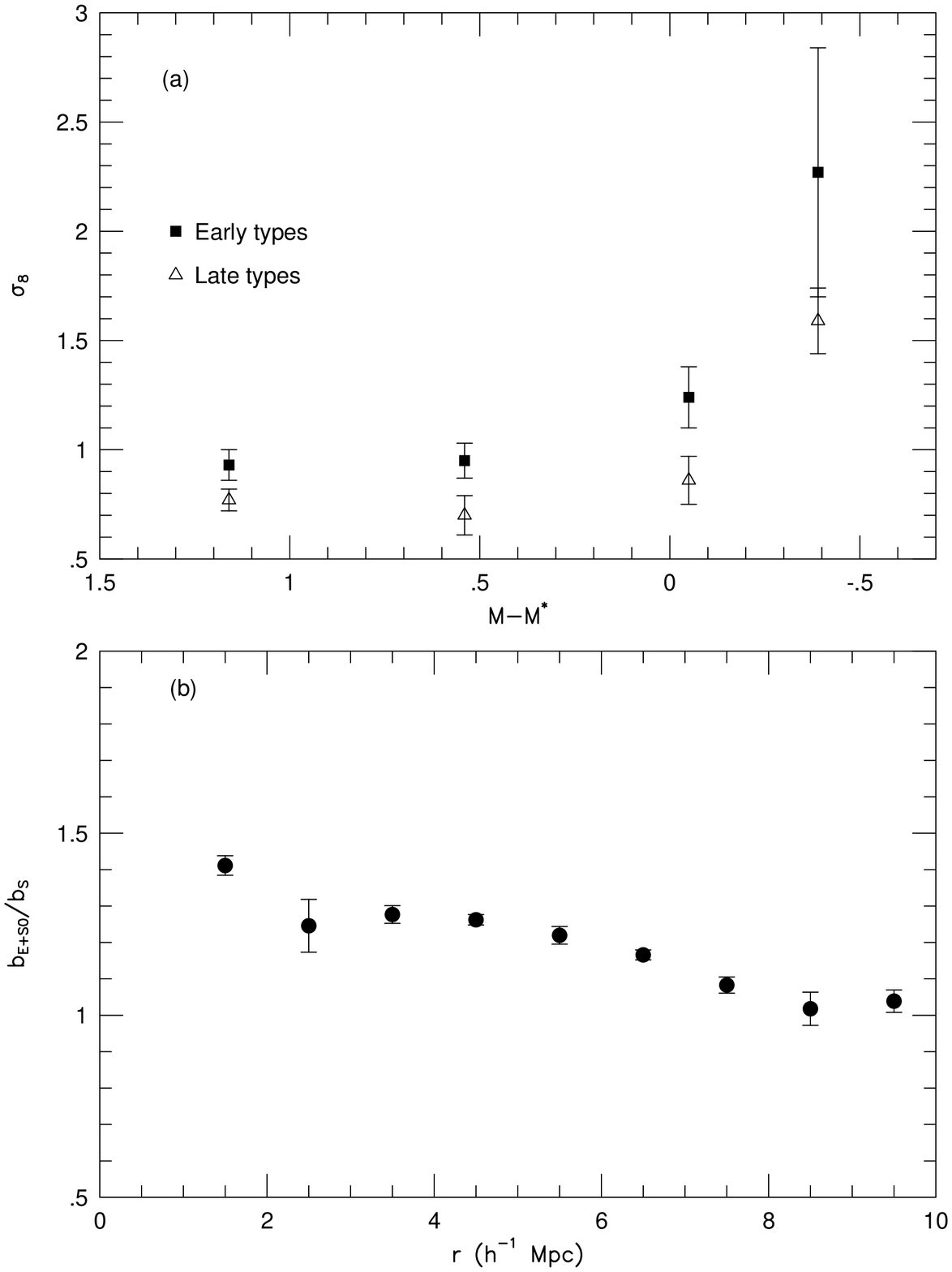,height=8cm}}
\vspace*{0.25cm}
\caption[]{Linear biasing measures for early/late-type galaxies. Panel
(a) shows the variance for different luminosity thresholds while panel
(b) shows the relative bias between early and late types as a function
of scale (for details see Willmer, da Costa \& Pellegrini 1998).}
\label{fig:biasing}
\end{figure*}

\subsection {Galaxy Properties and Biasing}

The large number of galaxies available in the nearby dense surveys has
made it possible to examine in greater detail the clustering
properties of galaxies of different types. Such studies may contribute
to our understanding of the relation between galaxies and LSS and help
constrain galaxy biasing models. Recent works based on the SSRS2
(Benoist \etal 1996, Willmer, da Costa \& Pellegrini 1998) have shown
evidence of strong, scale-independent luminosity bias, with more
luminous galaxies showing a much stronger correlation than sub-$L_*$
galaxies.  This result is in marked contrast with the findings based on
the Stromlo-APM survey (Loveday \etal 1995). The scale-independence
suggests that this bias may be established at the time of galaxy
formation. While several models of galaxy formation predict some
degree of luminosity bias (Mo \& White 1996), none can reproduce the
observed dependence on the luminosity. An interesting spin-off of this
analysis has been to find that very bright galaxies ($L \gsim 3L_*$)
show a large correlation length ($\sim$ 15 \h1 Mpc), comparable to
that observed for clusters (Cappi \etal 1998).  Interestingly, these
galaxies are not found preferentially in prominent association of
galaxies such as clusters or even loose groups. One possible
interpretation is that these galaxies may be associated with more
massive dark halos forming systems with atypically large $M/L$, which
would naturally account for their large correlation length.

Using the SSRS2, one also finds that the relative bias between early
and late types is scale-dependent (Figure~\ref{fig:biasing}), varying from about 1.4
on small scales to 1 at $\sim$ 8 \h1 Mpc, which may suggest that
environmental effects may play a role. The mean relative bias is found
to be $b_E/b_L$ $\sim$ 1.2. This small value, when compared to
previous surveys (\eg Guzzo \etal 1997), probably reflects the paucity
of rich clusters in the surveyed region.  Both early and late types
separately show a luminosity-dependent bias similar to the sample as a
whole further suggesting that the luminosity bias is primordial in
nature while the excess clustering of early types relative to spirals
on small scales may be caused by environmental effects.  The relative
bias between red and blue galaxies is similar to that observed between
early and late type galaxies. However, it levels off on smaller scales
$\sim$ 4 \h1 Mpc at a constant value of about 1.2.  The mean relative
bias of galaxies selected by colors is greater than when selected by
morphologies. It is important to emphasize that although galaxy
morphology and color are related, the scatter is large. This means
that these properties may be considered as independent characteristics
with colors reflecting the star formation history of galaxies. These
results are in qualitative agreement with theoretical predictions of
Kauffmann, Nusser \& Steinmetz (1997). Finally, one finds that the
relative bias between optical and \iras galaxies also varies with
scale at least out to $\sim$ 10 \h1 Mpc and shows a strong luminosity
dependence. The mean relative bias between optical and \iras is
$b_o/b_I \sim 1.5 $ in real space.

\begin{figure*}
\centering\mbox{\psfig{figure=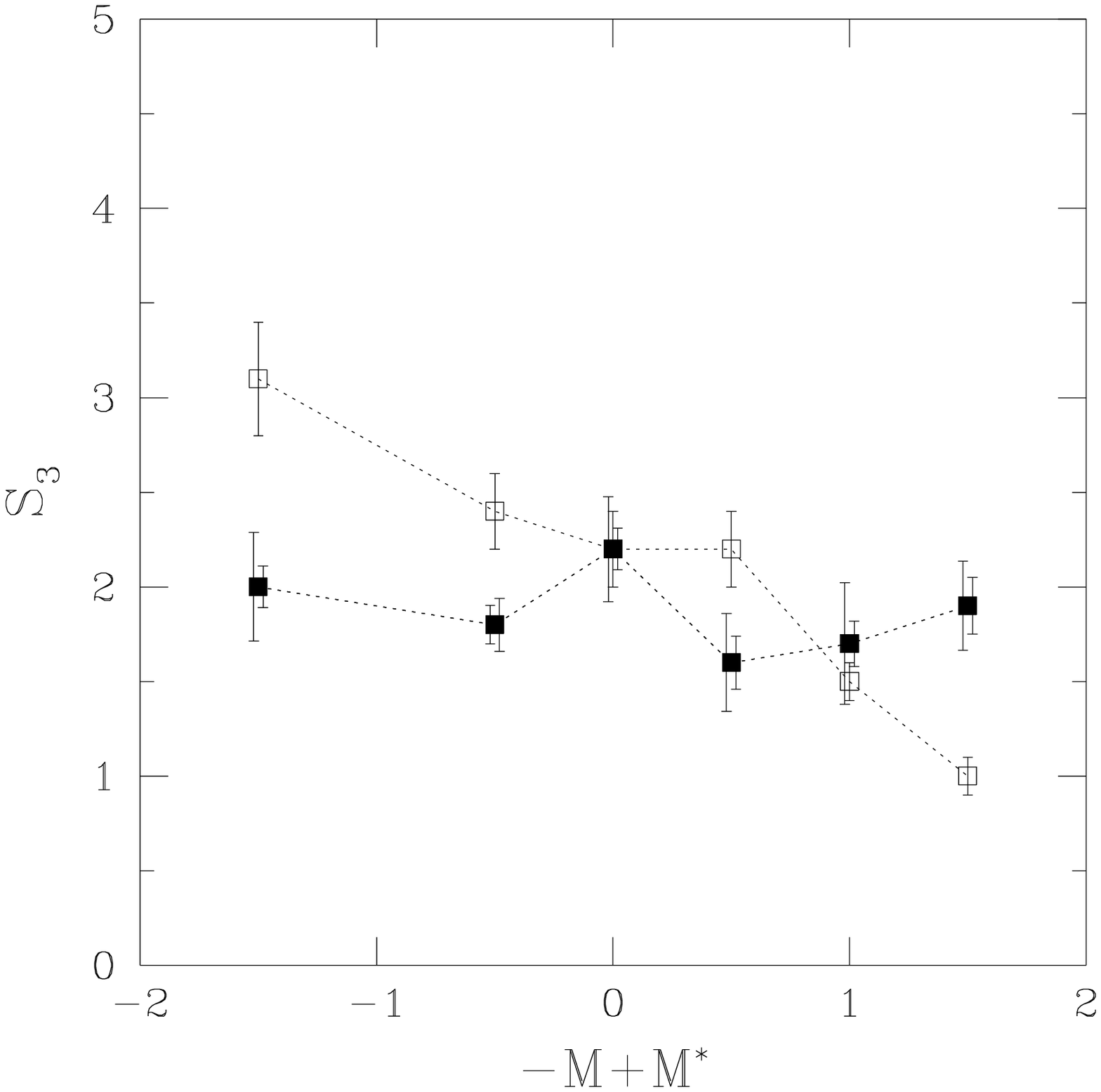,height=8cm}}
\vspace*{0.25cm}
\caption[]{The measured skewness $S_3$ of different volume-limited
  sub-samples (full squares) compared to the expected skewness in the
  linear bias scenario (open squares). Two estimates of the errors are
  displayed. For sake of clarity, they are slightly shifted in
  magnitude. The left error bars are estimated from mock
  volume-limited samples extracted from a standard CDM simulation
  having the same geometry of the SSRS2, and the right error bars are
  estimated from the bootstrap method (for details see Benoist \etal
  1998).}
\label{fig:s3}
\end{figure*}

Additional information on the nature of bias can be extracted by
investigating the higher order moments of the galaxy
distribution. This type of investigation has been conducted using the
the two-dimensional APM catalog (Gazt\~naga \& Frieman 1994) and in
three-dimensions by Bouchet \etal (1993) using the 1.2 Jy \iras
Catalog. More recently, Benoist \etal (1998) have carried out similar
analysis by comparing the measured skewness in volume-limited catalogs
extracted from the SSRS2. Using the large number of galaxies they
measured $S_3$ for different volume-limited samples finding $S_3$
scale- and luminosity-independent. As shown in Figure~\ref{fig:s3}, the weak
dependence of $S_3$ on luminosity is in marked contrast to what would
be expected from the strong dependence of the two-point correlation
function on luminosity in the framework of a linear biasing
model. This result seems to argue in favor of some degree of
non-linear bias.

The information derived from studies of clustering as a function of
the internal properties of galaxies such as luminosity,
color. morphology and internal dynamics, are essential for
understanding the connection between galaxy formation and LSS. The
present results are merely a preview of what will be possible with the
data from a complete redshift survey of a multicolor sample of
galaxies as envisioned by SDSS nearby and the ongoing work at
high-redshift.

\section{Summary}

The progress made by redshift surveys has been truly remarkable and
promises to continue to be so in the foreseeable future. We are now in
a curious transition period.  While some basic questions such as the
normalization and faint-end of local LF and the scale of the largest
inhomogeneities remain open, information about the clustering
properties of galaxies at $z\sim3$ are being studied. Clearly, it is
just a matter of time for a more definite picture of the galaxy
distribution and the time evolution of galaxy clustering to
emerge. Even though this may not yet provide a definite constraint to
the background cosmology it will certainly provide important data to
confront galaxy evolution models and answers to how, where, and when
galaxies formed. Even though the scientific goals may have changed, it
is clear that redshift surveys will continue to be an important
cosmological probe for the next 20 years.

\section*{Acknowledgments}
I would like to thank the Directors of MPA and ESO for making the
organization of this joint meeting possible and Tony Banday for
putting it altogether.


\end{document}